\documentclass[twocolumn]{jpsj2} 
\title{Incommensurate Lattice Distortion in the High Temperature 
Tetragonal Phase of La$_{2-x}$(Sr,Ba)$_{x}$CuO$_{4}$}

\author{
Shuichi \textsc{Wakimoto}$^{1}$
\thanks{E-mail address: wakimoto.shuichi@jaea.go.jp}, 
Hiroyuki \textsc{Kimura}$^{2}$,
Masaki \textsc{Fujita}$^{3}$,
Kazuyoshi \textsc{Yamada}$^{3}$,
Yukio \textsc{Noda}$^{2}$, \\
Gen \textsc{Shirane}$^{4}$,
Genda \textsc{Gu}$^{4}$,
Hyunkyung \textsc{Kim}$^{5}$,
and Robert J. \textsc{Birgeneau}$^{6}$
}

\inst{
$^{1}$Quantum Beam Science Directorate, Japan Atomic Energy Agency,
   Tokai, Ibaraki 319-1195, Japan \\
$^{2}$Institute of Multidisciplinary Research for Advanced Materials,   
   Tohoku University, Sendai 980-8577, Japan \\
$^{3}$Institute for Material Research, Tohoku University, 
   Katahira, Sendai 980-8577, Japan \\
$^{4}$Department of Physics, Brookhaven National Laboratory,
   Upton, New York 11973-5000, USA \\
$^{5}$Department of Physics, University of Toronto, Toronto,
   Ontario, Canada M5S~1A7 \\
$^{6}$Department of Physics, University of California, 
   Berkeley, Berkeley, California 94720-7300, USA
}

\abst{
We report incommensurate diffuse (ICD) scattering 
appearing in the high-temperature-tetragonal (HTT) phase 
of La$_{2-x}$(Sr,Ba)$_{x}$CuO$_{4}$ with $0.07 \leq x 
\leq 0.20$ observed by the neutron diffraction technique.  
For all compositions, a sharp superlattice peak of the 
low-temperature-orthorhombic (LTO) structure is replaced 
by a pair of ICD peaks with the modulation vector parallel to 
the CuO$_6$ octahedral tilting direction, that is,
the diagonal Cu-Cu direction of the CuO$_2$ plane, above the 
LTO-HTT transition temperature $T_s$.  The temperature 
dependences of the incommensurability $\delta$ for all 
samples scale approximately as $T/T_s$, while those of the
integrated intensity of the ICD peaks scale as $(T-T_s)^{-1}$.
These observations together with absence of ICD peaks in 
the non-superconducting $x=0.05$ sample evince a universal 
incommensurate lattice instability of hole-doped 214 
cuprates in the superconducting regime.
}

\kword{La$_{2-x}$Sr$_{x}$CuO$_{4}$, High-$T_{\rm c}$ superconductivity, 
Neutron scattering, Diffuse scattering}

\begin{document}
\maketitle

\section{Introduction} 

After the discovery of the high-transition-temperature 
(high-$T_{\rm c}$) superconductivity, extensive efforts have been 
made in this research field for almost two decades.  Although 
the mechanism of the Cooper pair formation in the high-$T_{\rm c}$ 
systems still requires more clues to be properly understood, 
it is widely expected 
that the incommensurate magnetic state is intimately 
related to the superconductivity.
In fact, neutron scattering experiments for the 
hole-doped superconductor La$_{2-x}$Sr$_{x}$CuO$_{4}$ have 
revealed incommensurate magnetic fluctuations~\cite
{Cheong91,Yamada_98} which exists up to the critical hole 
concentration where the superconductivity disappears.~\cite
{Wakimoto_OD} 
Such incommensurate magnetic fluctuations have been frequently 
ascribed to a dynamic microscopic phase separation into hole-free 
magnetic regions divided by hole-rich stripes,~\cite
{JMT_Nature95,Machida_89,Schulz_89,Zaanen_89,Emery_97} 
which may suggest that the high-$T_{\rm c}$ systems are 
intrinsically unstable to stripe-like formation on fundamental 
lattices.

By x-ray scattering techniques, incommensurate diffuse (ICD)
peaks  have been observed with modulation vectors along the 
orthorhombic $a$- and $b$-axes, that is, along the diagonal 
Cu-Cu direction of the CuO$_2$ plane, in the 
low-temperature-orthorhombic (LTO) phase of 
La$_{2-x}$Sr$_{x}$CuO$_{4}$ and 
La$_{2-x}$Sr$_{x}$NiO$_{4}$.~\cite{Isaacs_94,Dmowski_95}
Later, a coexistence of two types of CuO$_{6}$ octahedral 
tilt, one of the LTO type and the other of the 
low-temperature-tetragonal (LTT) type, have been reported by 
x-ray absorption fine structure (EXAFS) and x-ray 
absorption near edge structure (XANES) measurements although 
the average crystal structure is 
LTO.~\cite{Bianconi_96,Saini_97}  From 
these results it has been hypothesized that a local LTT 
structure exists in a stripe form in the LTO phase.

Similar to this, it has been reported that an LTO type 
octahedral tilt also remains locally in the 
high-temperature-tetragonal (HTT) phase 
in La$_{2-x}$Sr$_{x}$CuO$_{4}$ by 
EXAFS,~\cite{Haskel_96} nuclear magnetic resonance 
(NMR),~\cite{Goto_proc} and 
neutron diffraction.~\cite{Braden_01,Kimura_00}
By neutron measurements, Kimura {\it et 
al.}~\cite{Kimura_00} have observed a local LTO-type distortion 
in the HTT phase of La$_{2-x}$Sr$_{x}$CuO$_{4}$ with $x=0.12$ and 
$0.18$.  The evidence for this is incommensurate diffuse 
peaks at $((n \pm \delta)/2, (n \pm \delta)/2, m)$ ($n$:odd, 
$m$:even) as shown by the open circles in Fig.~\ref{fig01}(a).  
Note that in this expression the 
incommensurability $\delta$ is equivalent to the distance 
between the ICD peak and the LTO superlattice position expressed 
in orthorhombic reciprocal lattice units (r.l.u.).  This 
modulation direction corresponds to, again, the diagonal 
Cu-Cu direction.  Moreover, the incommensurability $\delta$ 
for the two concentrations scales approximately as the  
normalized temperature $T/T_s$ where $T_s$ is the LTO-HTT 
transition temperature.  Recently, it has been reported 
that no such incommensurate diffuse feature can be detectable in 
the non-superconducting $x=0.05$ sample,~\cite{Wakimoto_04} 
suggesting a correlation between the incommensurate structural 
instability and the superconductivity.

In the present paper, we report the observation of incommensurate 
diffuse scattering in the HTT phase studied by 
neutron diffraction using single crystals of 
La$_{2-x}$(Sr,Ba)$_{x}$CuO$_{4}$ with $x=0.07, 0.125, 0.15$ and 
$0.20$.  These samples cover a wide concentration range from the 
underdoped to slightly overdoped regions.  The intent 
of these measurements is to 
understand the relation between the structural instability and 
the superconductivity.

\section{Experimental Details}

Single crystals of La$_{2-x}$Sr$_{x}$CuO$_{4}$ (LSCO) 
with $x=0.07, 0.15$, and $0.20$ and 
La$_{2-x}$Ba$_{x}$CuO$_{4}$ (LBCO) with $x=0.125$ were 
grown by the traveling-solvent floating-zone 
method.~\cite{Hosoya94,CHLee_98}  All crystals underwent 
post-growth anneal in an oxygen atmosphere to remove any oxygen 
deficiencies.  The typical sample size was 7 mm in diameter and 
30 mm in length.  Neutron scattering experiments were 
performed at the TOPAN thermal neutron triple-axis 
spectrometer and the LTAS cold neutron triple axis 
spectrometer at the Japan Atomic Energy Agency in Tokai, Japan.  

ICD peaks of LSCO with 
$x=0.15$ and LBCO with $x=0.125$ were measured 
mainly at the (3/2, 3/2, 2) position using TOPAN with a 
collimation sequence of B-60$'$-60$'$-B and an incident 
neutron energy $E_i=13.5$~meV ($\lambda=2.46$~\AA), and 
those of LSCO with $x=0.07$ and $0.20$ 
were studied mainly at (1/2, 1/2, 4) using LTAS with 
G-80$'$-80$'$-B and $E_i=5$~meV ($\lambda=4.05$~\AA).  Pyrolytic 
graphite and Be filters were utilized for the TOPAN and the 
LTAS spectrometers, respectively, to remove neutrons with 
higher harmonic wave length $(\lambda/2, \lambda/3, 
etc.).$

\begin{figure}[tb]
\begin{center}
\includegraphics[width=85mm]{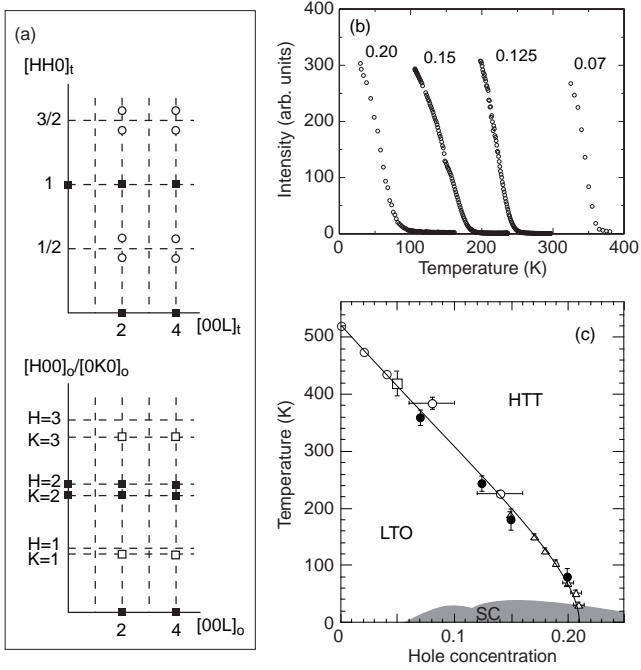}
\end{center}
\caption{(a) Neutron 
scattering peak geometries for $T>T_s$ (top) and $T<T_s$ 
(bottom).  The closed square, open squares, and open circles
represent the nuclear Bragg peaks, superlattice peaks in the 
LTO phase, and the incommensurate diffuse peaks, respectively.
(b) Temperature dependence of the superlattice 
peak intensity measured at the (014) position in the LTO notation.  
(c) Hole concentration dependence of the LTO-HTT transition 
temperature $T_s$.  The open symbols are data referred from 
refs~\citen{Wakimoto_04,Keimer_92,Takagi_92,Suzuki_98,
Kimura_99}, while the closed symbols represent $T_s$ of the 
present samples which are derived by a linear extrapolation 
of the temperature dependence of superlattice peak intensity 
to zero.}
\label{fig01}
\end{figure}

All samples show a structural transition from the HTT to the LTO 
structures.  Figure~\ref{fig01}(a) depicts the peak geometries in 
the reciprocal lattice above (top) and below (bottom) the 
transition temperature $T_s$.  In the LTO phase $(T<T_s)$, 
superlattice peaks appear at positions such as (014) 
and (032) shown by the open squares.  Note that in the LTO 
phase the orthorhombic $a$- and $b$-axes are superposed due 
to the twinned structure.  As temperature increases above 
$T_s$, the superlattice peak is replaced by weak  
incommensurate diffuse peaks as shown by the open circles
in Fig.~\ref{fig01}(a) (Top).
Figure~\ref{fig01}(b) shows the temperature dependence of the 
superlattice peak intensity for all samples measured at the 
(014) position.  
$T_s$ is determined to be 359~K, 241~K, 180~K, and 80~K for 
$x=0.07, 0.125, 0.15$, and 0.20, respectively, by a linear 
extrapolation of the superlattice peak intensity to zero.
Those values are summarized in Fig.~\ref{fig01}(c) 
together with those for different concentrations reported 
previously.~\cite{Wakimoto_04,Keimer_92,Takagi_92,Suzuki_98,Kimura_99} 
It is shown that the present crystals have Sr contents that vary 
systematically and are consistent with the 
nominal compositions. 

In the present paper, we utilize tetragonal 
notation to express the ${\rm\bf Q}$ positions of the 
incommensurate diffuse peaks in the HTT phase.  The tetragonal 
notation gives $a=3.8$~\AA~ and $c=13.1$~\AA, corresponding to 
reciprocal lattice units $a^{*}=1.65$~\AA$^{-1}$ and 
$c^{*}=0.47$~\AA$^{-1}$, while the orthorhombic reciprocal 
lattice unit is to $a^{*}/\sqrt 2 = 1.17$~\AA$^{-1}$.

\section{Results}

\begin{figure}[tb]
\begin{center}
\includegraphics[width=85mm]{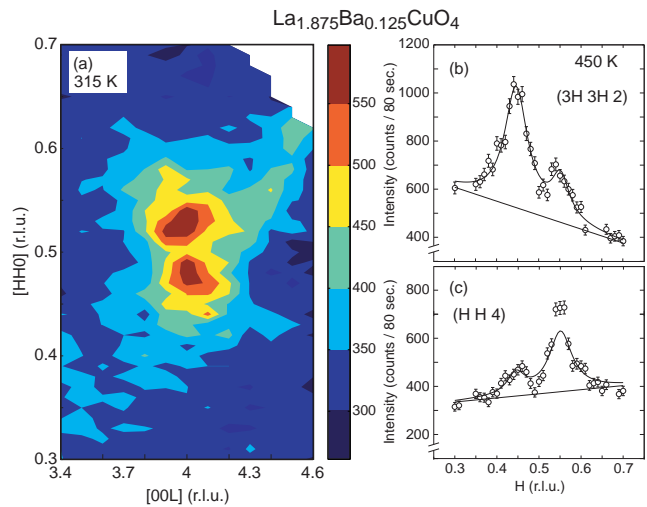}
\end{center}
\caption{(a) Contour plot of the incommensurate diffuse 
peaks around $(1/2, 1/2, 4)$ of La$_{1.875}$Ba$_
{0.125}$CuO$_4$ measured at 315~K.  (b) and (c) 
show incommensurate peak profiles around $(3/2, 3/2, 2)$ 
and $(1/2, 1/2, 4)$, respectively, measured at 450~K along 
the [HHO] direction.  The solid lines in (b) and (c) are the 
results of fits to a 
resolution-convoluted 2D Lorentzian function and 
adjusted background levels.}
\label{fig02}
\end{figure}

We have observed incommensurate diffuse peaks in the HTT phase 
for all of the present crystals.  Figure~\ref{fig02}(a) is a 
representative contour plot of the ICD peaks of 
LBCO $x=0.125$ measured at 315~K around the $(1/2, 1/2, 
4)$ position in the $HHL$ scattering plane.  As the 
ICD peaks are observed clearly in the LBCO sample, 
the ICD feature in 
the HTT phase appears to be common in the hole-doped 214 
compounds.  We have scanned along the [H-H0] 
direction at the ICD peak position by changing the tilt 
angle of the sample.  This scan indicates that the ICD peaks 
are located primarily on the $HHL$ plane, that is, the modulation 
direction is the [HH0] direction which corresponds to the 
diagonal Cu-Cu direction on the CuO$_2$ plane.  

To study the ${\rm\bf Q}$-dependence, we have measured the 
ICD peaks in different zones.  Figures~\ref{fig02}(b) and 
\ref{fig02}(c) show peak profiles at 450~K along the [HH0] 
direction at the (3/2, 3/2, 2) and (1/2, 1/2, 4) positions, 
respectively.
We note that there is an intensity imbalance between the ICD 
peaks, which increases as the temperature increases.  It is also 
notable that the intensity imbalance alters between the 
$(3/2, 3/2, 2)$ and $(1/2, 1/2, 4)$ zones, that is, the ICD 
peak at smaller $H$ is higher in the former zone whereas that 
at larger $H$ is higher in the latter.  Those features are 
commonly observed in all of the present samples.

The profiles in Figs.~\ref{fig02}(b) and \ref{fig02}(c) have been 
fitted to a resolution-convoluted two-dimensional (2D) Lorentzian 
function 
\begin{equation}
S({\rm\bf Q}) = |F({\rm\bf Q})|^{2}
[\frac{1}{1+({\rm\bf q}-{\rm\bf q_{\delta}})^2\xi^2} + 
\frac{\alpha}{1+({\rm\bf q}-{\rm\bf q_{-\delta}})^2\xi^2}],
\end{equation}
where $F({\rm\bf Q})$ is the structure factor, 
${\rm\bf q_{\pm\delta}}$ is the ICD peak positions, $\xi$ is the 
correlation length, and $\alpha$ is a factor to compensate for 
the intensity imbalance. 
The spectrum distribution along the energy transfer $(\omega)$ 
direction is assumed to be well-concentrated at $\omega=0$, 
since, as shown later, the ICD signal is consistently observed 
by using the cold neutrons which give finer energy resolution.  
The fits  so-obtained are shown by the solid curves in 
Figs.~\ref{fig02}(b) and \ref{fig02}(c), where the background 
levels have also been adjusted.

Figures~\ref{fig02}(b) and \ref{fig02}(c) show larger diffuse 
intensity in the $(3/2, 3/2, 2)$ zone.  From the fitting 
parameters, we find the ratio of $|F({\rm\bf Q})|^{2}(1+\alpha)$, 
which corresponds to the summation of the peak intensities, 
between $(3/2, 3/2, 2)$ and $(1/2, 1/2, 4)$ to be $1.8$.  This 
value is consistent with the ratio of $|F({\rm\bf Q})|^{2}$ of 
the LTO superlattice peaks at (032) and (014) in the orthorhombic 
notation, $|F(032)|^{2}/|F({014})|^{2}=1.96$, suggesting that the 
ICD peaks 
originate from the atomic displacements corresponding to the LTO 
type octahedral tilts.

\begin{figure}[tb]
\begin{center}
\includegraphics[width=85mm]{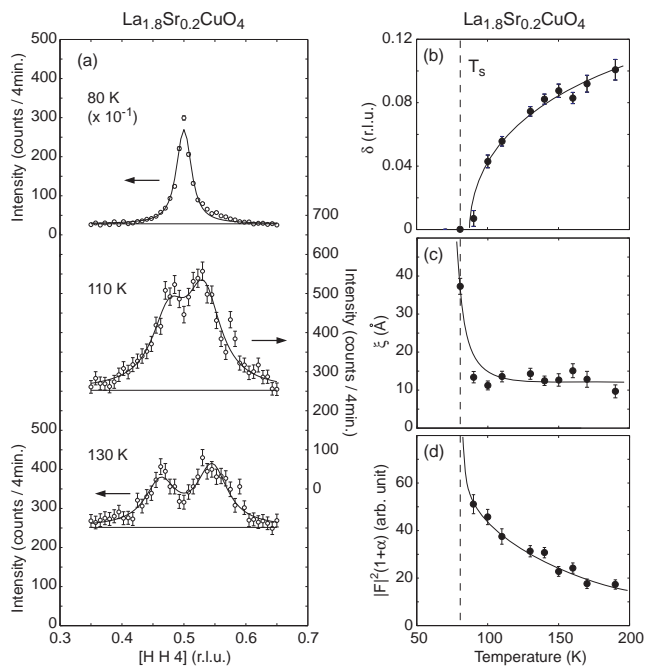}
\end{center}
\caption{(a) Temperature variation of the diffuse peak 
profile measured at $(1/2, 1/2, 4)$ for La$_{1.8}$Sr$_
{0.2}$CuO$_{4}$.  The solid lines are fits to a 
resolution-convoluted 2D Lorentzian function and 
adjusted background levels.  (b), (c), and 
(d) show temperature dependences of the incommensurability 
$\delta$, the correlation length $\xi$, and $|F|^2(1+
\alpha)$, respectively, derived from the fitting results.  
Solid lines are guides to the eye.}
\label{fig03}
\end{figure}

We have studied the temperature dependence of the ICD peaks.  
Figure~\ref{fig03}(a) shows the temperature dependence of the ICD 
peaks of the LSCO $x=0.20$ sample around $(1/2, 1/2, 4)$ 
measured along the $[HH0]$ direction.  Note that these 
profiles are taken by using cold neutrons with 
$E_i=5$~meV, while the data in Fig.~\ref{fig02} are taken by using 
thermal neutrons with $E_i=13.5$~meV.  We have observed 
qualitatively similar behavior of the ICD peaks for all of the 
samples in regardless of the neutron energy.  
We show the data of LSCO $x=0.20$ 
representatively to report the temperature dependences of 
physical quantities since the $x=0.20$ sample has the lowest $T_s$ 
among the samples; 
hence, we are able to study the ICD peaks up to reasonably 
high temperatures with respect to $T_s$. 
Figure~\ref{fig03}(a) clearly demonstrates that a commensurate 
peak near the transition temperature $T_s$ is replaced by 
incommensurate diffuse peaks at higher temperatures with an 
incommensurability that increases with increasing temperature.  

To draw more details, we analyze the data by fitting to the 
resolution-convoluted 2D Lorentzian function described in eq. (1).  
The results of fits are shown by the solid lines in 
Fig.~\ref{fig03}(a) and the parameters are summarized in 
Figs.~\ref{fig03}(b), \ref{fig03}(c) and \ref{fig03}(d).  The 
incommensurability $\delta$ decreases as temperature 
decreases and reaches zero near $T_s$, while the 
correlation length $\xi$ stays constant in the 
incommensurate region and diverges near $T_s$.  The 
quantity $|F|^2(1+\alpha)$ which corresponds to the 
summation of the peak intensities increases gradually as 
temperature decreases and shows a rapid increase near $T_s$.  
These behaviors of the fitting parameters are commonly observed 
for all samples.  Interestingly, the deconvolution analyses 
exhibit a constant correlation length of $\sim 15$~\AA~ above 
$T_s$ for all samples.  We show more details of the 
incommensurability and the integrated intensity later.
Also, a possible microscopic picture of these phenomena is 
discussed in the next section.

\begin{figure}[tb]
\begin{center}
\includegraphics[width=85mm]{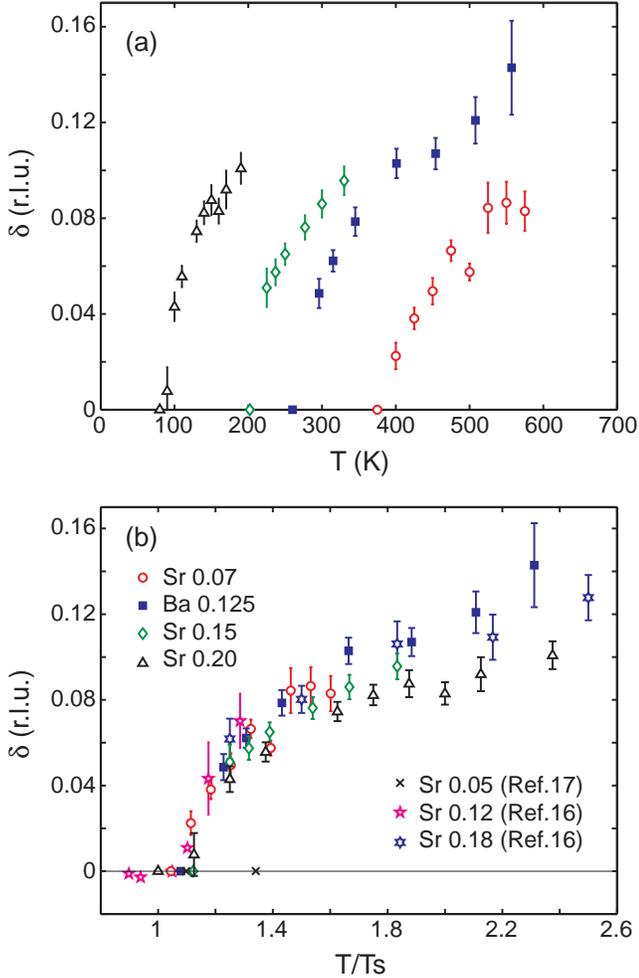}
\end{center}
\caption{(Color online) Incommensurability $\delta$ for all 
of the present samples as a function of (a) temperature and 
(b) normalized temperature $T/T_s$.  In (b) data of Sr $x=0.12$ 
and $0.18$ form ref.~\citen{Kimura_00}, and of $x=0.05$ 
from ref.~\citen{Wakimoto_04} are also shown.}
\label{fig04}
\end{figure}

Kimura {\it et al.}~\cite{Kimura_00} have reported that the 
temperature dependences of $\delta$ for the LSCO $x=0.12$ 
and $0.18$ samples fall onto an identical line  by 
utilizing a normalized temperature $T/T_s$.  We have tested 
if this applies to a wider concentration range.  
Figure~\ref{fig04}(a) shows the temperature dependence of the 
incommensurability $\delta$ for all samples.  It is found that, 
particularly by comparison between the data of $x=0.20$ and 
$0.07$, the gradient of the increase of $\delta$ is apparently 
smaller for samples with higher $T_s$.  Such behavior is 
consistent with the scaling feature of $\delta$ by $T/T_s$.  
In fact, $\delta$ approximately falls onto a universal curve 
as a function of $T/T_s$ as shown in Fig.~\ref{fig04}(b).  
Although, at high $T/T_s (>1.8)$, there are discrepancies 
between the data of $x=0.125$ and $0.20$, $\delta(T/T_s)$ 
agrees for the temperature range of $1 \leq T/T_s \leq 1.5$.
Since the universality holds for data measured by both 
thermal and cold neutrons, i.e. different energy 
resolutions, we conclude that the ICD feature is 
static and intrinsic rather than being caused by a spurious 
process, such as detecting phonons at the edge of 
resolution ellipsoid.  

As reviewed in the introduction, the ICD peaks were absent
for the LSCO $x=0.05$ sample.~\cite{Wakimoto_04}  
For this sample it has been 
confirmed that there is a weak commensurate diffuse peak up to 
550~K ($T/T_s \sim 1.34$).  Data for $x=0.05$ are plotted by 
crosses in Fig.~\ref{fig04}(b). 
Remarkably, the diffuse peak of the non-superconducting $x=0.05$ 
sample stays commensurate at $T/T_s=1.34$ whereas the present 
superconducting samples show an incommensurability of $0.06$ r.l.u. 
at the reduced temperature.  
This demonstrates that the ICD feature is characteristic to the 
superconducting samples.

\begin{figure}[tb]
\begin{center}
\includegraphics[width=85mm]{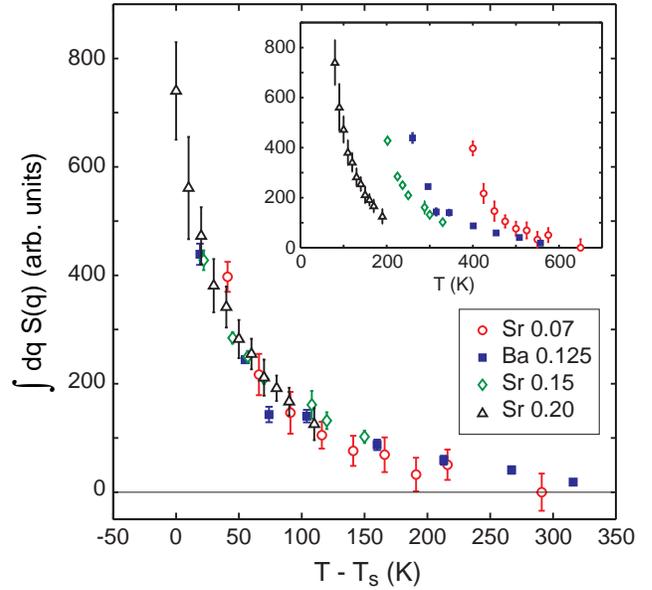}
\end{center}
\caption{(Color online) Normalized integrated intensity 
$\int dq S(q)$ for all of the present samples as a function 
of $T-T_s$.  The inset shows same quantities as a function 
of temperature.}
\label{fig05}
\end{figure}

To test if there is any unusual enhancement of the ICD 
peaks at a certain hole concentration, we compared integrated 
intensities, which have been derived by fitting the profiles 
to a simple Gaussian function and integrating the intensity 
along the [HH0] scan direction.  The derived integrated 
intensities have been normalized to sample volume, 
resolution volume, and structure factors.  Sample volume and 
resolution volume are determined by acoustic phonon 
measurements around $(1, 1, 0)$, while the structure factor 
ratio of 1.8 between $(3/2, 3/2, 2)$ and $(1/4, 1/4, 4)$ is 
adopted from the data of $x=0.125$ in Figs.~\ref{fig02}(b) 
and \ref{fig02}(c).  Note that, although the normalization to the 
resolution volume may have some uncertainty, each comparison 
between the samples of $x=0.07$ and $0.20$, and between the 
samples of $x=0.125$ and $0.15$ is robust since each pair has 
been measured under an identical spectrometer configuration, 
respectively.
The normalized integrated intensities, $\int dq S(q)$, so-obtained 
are summarized in Fig.~\ref{fig05} as a function of $T-T_s$.  
The inset shows the same quantities as a function of 
temperature.  The excellent agreement of $\int dq S(q)$ for 
all samples indicates that there is no particular 
enhancement of the ICD peak at a certain hole concentration.

\section{Discussion}

We have shown that the hole-doped 214 cuprates have 
universal incommensurate diffuse scattering in the HTT 
phase. 
From the comparison to the non-superconducting $x=0.05$ 
sample in Fig.~\ref{fig04}, it is very likely that the 
ICD feature is unique to the superconducting samples.  
It is also important to study superconducting samples 
without structural transition, such as overdoped samples.  
Our preliminary measurements of the superconducting 
$x=0.25$ sample at 12~K appear to exhibit very weak ICD 
peaks with $\delta \sim 0.1$ r.l.u.  
Thus, we believe that there is a universal local 
lattice instability towards the incommensurate octahedral 
tilt distortion in the superconducting samples.  If the system 
has a structural transition, the incommensurability is 
suppressed following the universal curve in 
Fig.~\ref{fig04} at the temperature range of $T/T_s \leq 1.5$.

Bianconi {\it et al.}~\cite{Bianconi_96} hypothesized that a 
LTT-type local lattice distortion exists in a stripe form in 
the LTO background to explain the incommensurate diffuse scattering in 
the LTO phase.  A similar explanation appears to be relevant to 
the present observations of the ICD peaks in the HTT phase; that is, 
LTO-type octahedral tilts remain locally in the HTT phase 
with a one-dimensional modulation.  

\begin{figure}[tb]
\begin{center}
\includegraphics[width=85mm]{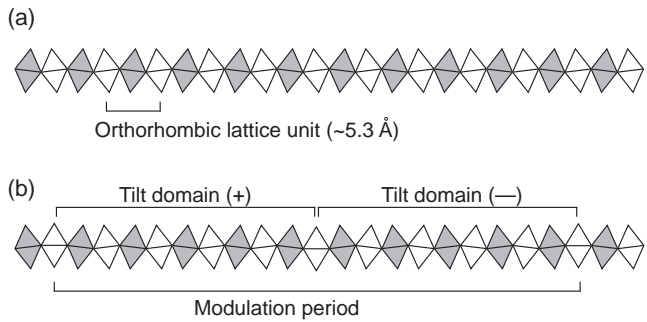}
\end{center}
\caption{(a) CuO$_6$ octahedral tilts in the LTO phase viewed 
from the orthorhombic $a$-axis.  (b) An example of modulated 
tilting pattern which causes incommensurate peaks.  Antiphased 
tilting domains are recognized as domain (+) and (-).  Depicted 
modulation period is 10 orthorhombic lattice units giving the 
incommensurability of 0.10 r.l.u.}
\label{fig06}
\end{figure}

As shown in Fig.~\ref{fig01}(a), the observed ICD peaks split 
in the $HHL$ scattering plane.  This means that the modulation 
vector is parallel to the LTO-type octahedral tilt-direction.  
Note that if the modulation vector is perpendicular to the tilting 
direction, the ICD peaks should split from the (3/2, 3/2, 2) 
and (1/2, 1/2, 4) positions along the perpendicular direction to 
the scattering plane. 

An example of the modulated octahedral tilting pattern that 
gives the observed incommensurate geometry is depicted in 
Fig.~\ref{fig06}(b), 
while Fig.~\ref{fig06}(a) shows uniform octahedral tilts, 
that give commensurate superlattice peak, viewed from the 
direction perpendicular to the tilting direction.  The 
modulation period depicted in Fig.~\ref{fig06}(b) is 10 orthorhombic 
lattice units, consisting of 5-lattice each of antiphase 
tilting domains, indicated by domains (+) and (-), which results 
in an incommensurability of 0.1 r.l.u.  Those domains may form 
stripes running perpendicular to the tilting direction.  
Alternatively, a 
tilting pattern with a varying tilt-angle instead of the 
stacking of antiphase tilting domains gives qualitatively 
the same incommensurate geometry.  The neutron scattering 
technique is, however, unable to distinguish between these patterns.  
If we assume a uniform tilt-angle of the octahedra in a single 
tilting domain as depicted in Fig.~\ref{fig06}(b), the  
modulation period observed to be more than 50~\AA~ 
is much longer than the observed correlation 
length of $\sim 15$~\AA.  Thus it might be natural to consider a 
varying tilt-angle with modulation.  Either model, however, does 
not affect the present discussions.

From the temperature dependence of the parameters in 
Fig.~\ref{fig03}, a possible microscopic picture can be 
suggested as follows.  At very high temperature $T/T_s \sim 
2.5$, local lattice distortions of the LTO structure start 
to appear in a one-dimensional form aligned parallel to the 
diagonal Cu-Cu direction.  At this point, each LTO stripe is thin 
and close to each other resulting in a small modulation period and 
large incommensurability.  As the temperature decreases, the LTO 
stripes grow thicker by merging with each other, resulting in 
larger modulation period and smaller incommensurability.  
In this growing process, the striped phase has a constant 
correlation length $\sim 15$~\AA.  Either by growing the volume 
of the striped phase or by enlarging the octahedral tilt-angle, 
the quantity of $|F|^2(1+\alpha)$ and integrated intensity 
increases.  Finally, the modulated phase evolves into a uniform 
tilt phase, and the correlation length diverges at $T_s$.

We note that, with decreasing temperature, the diffuse scattering 
becomes commensurate at $T/T_s \sim 1.15$ whereas the system show 
the structural phase transition at the lower temperature $T/T_s = 1$.  
The difference $T/T_s = 0.15$ corresponds to 10 to 30~K in actual 
temperature scale depending on $T_s$.  We infer that there is a 
crossover from the incommensurate to the commensurate states at 
$T/T_s \sim 1.15$ where the the size of the tilt domain, i.e., 
the coherence length of the in-phase tilting, far exceeds the 
correlation length scale which is nearly constant at $\sim 15$~\AA.  
Then the structural phase transition takes place at lower temperature 
where the correlation length diverges.

Although the model of local LTO distortion in a one-dimensional 
(stripe) form explains qualitatively the present observations, a few 
features still need to be clarified.  As mentioned in the previous 
section, a simple structure factor calculation can reproduce the ratio 
of the integrated intensity between (3/2, 3/2, 2) and (1/2, 1/2, 4).  
The change of imbalance, however, of the ICD peaks between these 
positions cannot be explained by the structure factor calculations 
based on the tilting pattern depicted in Fig.~\ref{fig06}(b).  
From the structure factor definition, the imbalance of the ICD 
peaks, i.e. the dependence of ICD intensity along the [HH0] direction, 
should originate from atomic displacements parallel to the 
tilting directions.
In the coordinates of the CuO$_6$ octahedral tilts, only apical 
oxygen atoms displace towards the tilting direction.  However, 
the structure factors of striped LTO-type 
tilts predicts larger ICD intensities at the 
higher $H$ position at both (3/2, 3/2, 2) and (1/2, 1/2, 4).
Thus, the unexpected 
change of imbalance at different zones may imply additional 
displacements of in-plane oxygen atoms along the tilt direction.
In this aspect, it is interesting to note possible correlations 
between the present ICD feature and the anomalous softening of 
Cu-O bond stretching mode observed by x-ray 
measurements.~\cite{Fukuda_05}  Both phenomena are observed in 
the superconducting regime, and, moreover, the latter feature may 
cause anomalous displacements of in-plane atoms.
More detailed measurements of the local 
structure are necessary to draw firm conclusions.

It is still an open question whether such incommensurate 
lattice distortion correlates directly with the formation of 
the magnetic incommensurate state at much lower temperatures.  
Kimura {\it et al.}~\cite{Kimura_00} have 
reported that the incommensurability of the ICD peaks 
saturates at about $0.12$ r.l.u. at high temperatures which value 
is close to the incommensurability of the magnetic state at low 
temperatures and thus they claimed an incipient lattice 
distortion starting at high temperatures.  
Present result of the incommensurability as a 
function of $T/T_s$ for a wide concentration range suggests 
a gradual increase of $\delta$ rather than saturation at high 
temperatures.  Moreover, the incommensurate modulation 
direction of the lattice distortion in the HTT phase of the 
present results and in the LTO phase observed by the x-ray 
measurements is consistently parallel to the diagonal Cu-Cu 
direction which is 45$^{\circ}$ away from the  magnetic 
incommensurate modulation direction.  Therefore it is 
unlikely that the lattice distortion is the direct origin of 
the charge and spin stripes.

Although not being the direct origin of the putative charge and 
magnetic stripes, the local lattice distortion can affect the charge 
stripes as a pinning potential.  In the hole-doped 214 cuprates, 
magnetic stripe order is achieved in both a LTO phase and a LTT 
phase.  Neutron scattering measurements report that the magnetic 
incommensurate peaks in the reciprocal space form rectangles 
elongated along the orthorhombic $b$-axis in the LTO 
phase,~\cite{Lee_99,Kimura_PRB00} while complete squares are 
observed in the LTT phase.~\cite{Tra_PRB96,Fujita_PRB04}  Since 
the LTT-type distortion pins the charge stripes more efficiently, 
entire stripes will be pinned in the uniform LTT phase, giving a 
square 
geometry of the IC peaks.  On the other hand, by assuming a LTT 
distortion in a stripe form running along the orthorhombic 
$b$-axis in an LTO background, the vertical charge stripes 
are pinned only at the intersections with the diagonal 
LTT stripes.  In this case, the charge stripes may order 
in a smectic form, which results in a rectangular geometry 
of the IC magnetic peaks.  Thus, the hypothesis of the lattice 
distortion in a stripe form appears to be consistent with 
the observations of the magnetic stripes.

In summary, we have studied incommensurate diffuse peaks 
appearing around the LTO superlattice positions with 
modulation vector along the diagonal Cu-Cu direction in the 
HTT phase of the hole-doped 214 cuprates.  By comparison 
of the structure factors at different zones, 
we conclude that the ICD peaks 
originate from LTO type displacements, which suggests that 
a LTO type local lattice distortion remains in a 
one-dimensionally modulated form 
in the HTT phase.

\section*{Acknowledgment}

The authors thank 
T. Goto, A. Oosawa, K. Kakurai, S. Larochelle, C.-H. Lee, M. 
Matsuda, and N. Metoki for invaluable discussions.  
The present work was supported by the US-Japan Cooperative 
Research Program on Neutron Scattering.  
Work at the University of Toronto is part of the Canadian 
Institute of Advanced Research and supported by the 
Natural Science and Engineering Research Council of Canada.  
Research at Tohoku University is supported by a Grant-In-Aid 
for Young Scientists B (13740216 and 15740194) and a 
Grant-in-Aid from the Japanese Ministry of Education, Culture, 
Sports, Science and Technology, while work at BNL is supported 
by the U. S. DOE under contact No. DE-AC02-98CH10886.  
The work at Lawrence Berkeley Laboratory is supported by the 
Office of Basic Energy Sciences, U.S. Department of Energy under 
contract number : DE-AC03-76SF00098.

\end{document}